# A very energetic supernova associated with the γ-ray burst of 29 March 2003

Jens Hjorth*, Jesper Sollerman†, Palle Møller‡, Johan P. U. Fynbo§, Stan E. Woosley‖,

Chryssa Kouveliotou¶, Nial R. Tanvir#, Jochen Greiner☆, Michael I. Andersen**,

Alberto J. Castro-Tirado††, José María Castro Cerón‡‡, Andrew S. Fruchter‡‡,

Javier Gorosabel‡‡,††, Páll Jakobsson*, Lex Kaper§§, Sylvio Klose‖‖,

Nicola Masetti¶¶, Holger Pedersen*, Kristian Pedersen*, Elena Pian##,

Eliana Palazzi¶¶, James E. Rhoads‡‡, Evert Rol§§, Edward P. J. van den Heuvel§§,

Paul M. Vreeswijk☆☆, Darach Watson*, Ralph A. M. J. Wijers§§

*Astronomical Observatory, NBIfAFG, University of Copenhagen, Juliane Maries Vej,
DK-2100 Copenhagen Ø, Denmark*

*† Stockholm Observatory, Department of Astronomy, AlbaNova, S-106 91 Stockholm,
Sweden*

*‡ European Southern Observatory, Karl-Schwarzschild-Strasse 2, D-85748 Garching
bei München, Germany*

*§ Department of Physics and Astronomy, University of Aarhus, Ny Munkegade,
DK-8000 Århus C, Denmark*

*‖ Department of Astronomy and Astrophysics, University of California, Santa Cruz,
California 95064, USA*

*¶ NSSTC, SD-50, 320 Sparkman Drive, Huntsville, Alabama 35805, USA*

*# Department of Physical Sciences, University of Hertfordshire, College Lane,
Hatfield, AL10 9AB, UK*

*☆ Max-Planck-Institut für extraterrestrische Physik, D-85741 Garching, Germany*

*\*\* Astrophysikalisches Institut, D-14482 Potsdam, Germany*

*†† Instituto de Astrofísica de Andalucía (IAA-CSIC), PO Box 03004, E-18080 Granada, Spain*

*‡‡ Space Telescope Science Institute, 3700 San Martin Drive, Baltimore, Maryland 21218, USA*

*§§ Astronomical Institute Anton Pannekoek, Kruislaan 403, NL-1098 SJ Amsterdam, Netherlands*

*‖‖ Thüringer Landessternwarte Tautenburg, D-07778 Tautenburg, Germany*

*¶¶ Istituto di Astrofisica Spaziale e Fisica Cosmica - Sezione di Bologna, CNR, via Gobetti 101, I-40129, Bologna, Italy*

*## INAF, Osservatorio Astronomico di Trieste, Via G.B. Tiepolo 11, I-34131 Trieste, Italy*

*☆☆ European Southern Observatory, Casilla 19001, Santiago 19, Chile*

**Over the past five years evidence has mounted that long-duration (> 2 s) γ-ray bursts (GRBs) —the most brilliant of all astronomical explosions—signal the collapse of massive stars in our Universe. This evidence was originally based on the probable association of one unusual GRB with a supernova[1], but now includes the association of GRBs with regions of massive star formation in distant galaxies[2,3], the appearance of supernova-like 'bumps' in the optical afterglow light curves of several bursts[4-6] and lines of freshly synthesized elements in the spectra of a few X-ray afterglows[7]. These observations support, but do not yet conclusively demonstrate, the idea that long-duration GRBs are associated with the deaths of massive stars, presumably arising from core collapse. Here we report evidence that**

**a very energetic supernova (a hypernova) was temporally and spatially coincident with a GRB at redshift $z = 0.1685$. The timing of the supernova indicates that it exploded within a few days of the GRB, strongly suggesting that core-collapse events can give rise to GRBs, thereby favouring the 'collapsar' model[8,9].**

The first GRB associated with an observed supernova event, SN1998bw, occurred on 25 April 1998 at a redshift of only $z = 0.0085$ (about 37 Mpc)[1]. GRB980425 remains by far the closest burst recorded to date. Its proximity and unusually low γ-ray energy budget (total equivalent isotropic energy release of ~$8 \times 10^{47}$ erg, about four orders of magnitude less than other long-duration bursts with measured redshifts) have led to some doubt that the progenitor of this GRB was the same as those of GRBs at high redshift. SN1998bw was itself unusual; it displayed kinetic energies more than an order of magnitude greater than typical Ic supernovae and a very high radio flux[10]. The tantalizing evidence for the association between GRB980425 and SN1998bw launched a relentless hunt for the 'smoking gun' signature of a supernova in the optical afterglow data of many subsequent GRBs.

On 29 March 2003 (11:37:14.67 UT) NASA's High Energy Transient Explorer, HETE-II, detected a very bright GRB[11]. GRB030329 lasted about 25 seconds and had a fluence of ~$1.2 \times 10^{-4}$ erg cm$^{-2}$ and a peak flux of ~$7 \times 10^{-6}$ erg cm$^{-2}$ s$^{-1}$ (30–400 keV band). The fluence alone places GRB030329 in the top 0.2% of the 2704 GRBs detected with the Burst And Transient Source Experiment (BATSE) during its nine years of operation. Rapid follow-up observations[12,13], within 1.5 hours, discovered an optical afterglow of magnitude ~12 in the R-band, making it brighter than any previously detected afterglow at the same time after burst. A very bright afterglow was also detected at other wavelengths, ranging from X-rays to radio (see ref. 12 and references therein).

Our collaboration[14] determined the redshift of GRB030329, using the UV-Visual Echelle Spectrograph (UVES) on the Very Large Telescope (VLT) at the European Southern Observatory (ESO), to be $z = 0.1685$; the second closest long-duration GRB ever studied, after GRB980425. Assuming a flat $\Omega_\Lambda = 0.7$ and $H_0 = 70\,\mathrm{km\,s^{-1}\,Mpc^{-1}}$ cosmology, GRB030329 was at a luminosity distance of 810 Mpc. At this distance, GRB030329 is typical in its γ-ray budget: its total isotropic energy release is ~$9 \times 10^{51}$ erg (30–400 keV band) and thus qualifies as a classical cosmological GRB. Its relative proximity also accounts for the extreme initial brightness of its optical counterpart[12]. Extrapolating the early afterglow power-law decline of the transient and allowing for a supernova light curve similar to that of SN1998bw, we expected to see a supernova emerge 10–20 days after the burst. We therefore conducted spectroscopic observations between 5 and 33 days after the GRB (Table 1), and searched in the spectra of the fading optical afterglow for an emerging supernova contribution.

Figure 1 displays the sequence of six spectra obtained with the VLT FOcal Reducer/low dispersion Spectrographs (FORS1 and FORS2) at the epochs listed in Table 1. At the first epoch (3 April), the GRB afterglow spectrum is reasonably well fitted by a single power law with a spectral-index $\beta \sim -1.2 \pm 0.1$ (where $f_\nu \propto \nu^{\beta}$), typical for GRB afterglows. By 8 April, it is clear that while the same underlying power-law is present, the small residuals visible on 3 April are now highly significant and resemble the spectrum of a supernova. The absence of broad hydrogen lines indicates a Type I supernova, and the weak or absent Si II $\lambda = 6,150$ Å ($\lambda6,150$) and He lines rule out Types Ia and Ib, respectively. The unusually broad features and blended lines in the spectrum therefore indicate a Type Ic supernova with a very large expansion velocity, remarkably similar to SN1998bw[15]. This supernova has been designated SN2003dh following the independent parallel discovery reported in refs 16 and 17.

To monitor the spectral evolution as well as the overall luminosity of SN2003dh, we decomposed the spectra into supernova and afterglow components, using as templates known spectra of SN1998bw at various stages of its evolution. In Fig. 2 we display the spectra of SN2003dh along with those of SN1998bw at various epochs. The decomposed spectra clearly bring out the emergent supernova features. At 8–10 days (restframe) after the burst, broad lines around 5,000 Å and 5,800 Å (restframe 4,300 Å and 5,000 Å) develop, resembling the SN1998bw spectrum[15]. After 28 days new lines emerge at 5,300 Å and 7,500 Å (restframe 4,550 Å and 6,400 Å), presumably due to Mg I] $\lambda$4,571 and [O I] $\lambda$6,300 and $\lambda$6,364, characteristic of Type Ic supernova spectra[15].

From the spectra we measure the expansion velocity ten days after GRB030329 from the minimum of the absorption trough of the Si II $\lambda$6,355 line to be 36,000 ± 3,000 km s[-1] (0.12 ± 0.01 *c*). This is considerably larger than any previously known supernova (for SN1998bw it was ~23,000 ± 3,000 km s[-1] ten days after the explosion[15]). Thus SN2003dh qualifies as the most extreme member of the class of peculiar supernovae colloquially known as 'hypernovae': supernovae with very broad lines indicative of expansion velocities in excess of ~30,000 km s[-1] at early times[18].

We have attempted to date the supernova explosion of SN2003dh spectroscopically, under the assumption of its parallel spectral evolution with SN1998bw. This assumption introduces a potential systematic uncertainty, but we can qualitatively conclude that SN2003dh began ± 2 days relative to GRB030329 (to this should be added the model dependent uncertainty of +0.7/–2 days of the SN1998bw dating[19]). A more detailed analysis will be presented elsewhere (J.S. *et al*., manuscript in preparation).

As an additional dating check, we have compared the V-band light curves of SN2003dh and 1998bw (Fig. 3a). SN2003dh peaked about 10–13 days (restframe) after GRB030329 whereas SN1998bw peaked about 17 days after GRB980425. At maximum, SN2003dh was slightly brighter than SN1998bw, consistent with its larger expansion velocity. The difference between the light curves indicates that SN2003dh may have preceded its associated GRB by about 4 – 7 days, somewhat earlier than in the case of SN1998bw. However, given their parallel spectral evolution (Fig. 2) we consider it more likely that the two supernovae were co-eval but displayed different light curves, as indicated by the differences in expansion velocity (Fig. 3b), peak brightness, detailed spectral evolution and strength of associated GRB.

We now consider the properties of SNe 2003dh and 1998bw in more detail. Both belong to the unusual class of Type Ic supernovae. SN1998bw was well explained by models involving the very energetic explosion of a massive Wolf–Rayet star[19,20] (a highly evolved star that has lost its outer hydrogen layers to a wind or companion star). Since Type Ic supernovae have light curves powered by radioactive decay, the bright light curve indicates the synthesis of at least several tenths of a solar mass of $^{56}$Ni in both events (ref. 21 and the current paper). Modelled in spherical symmetry—which is certainly a questionable hypothesis—SN1998bw required $1$–$2 \times 10^{52}$ erg of kinetic energy[19,20]. The energy requirements here are similar. However, it is much more likely that the explosion was asymmetric; in fact the rapid rise and decay of SN2003dh seen in Fig. 3 may indicate that we are viewing an asymmetric supernova along its axis of most rapid expansion[9,20]. The necessary energy may therefore be considerably smaller[20]. Further, the larger ratio of afterglow luminosity to supernova luminosity in GRB030329 is also consistent with a burst that has been seen nearly pole-on[22], whereas the low luminosity of GRB980425 and its afterglow may have been due to its having being viewed substantially off-axis, or simply because it ejected less highly relativistic matter.

The combined results on SNe 1998bw and 2003dh offer the most direct evidence yet that typical, long-duration, energetic GRBs result from the deaths of massive stars. The lack of hydrogen lines in both spectra is consistent with model expectations that the star lost its hydrogen envelope to become a Wolf–Rayet star before exploding. The broad lines are also suggestive of an asymmetric explosion viewed along the axis of most rapid expansion. The large $^{56}$Ni abundance needed for the light curve cannot have been made by the low-density jet; its mass and solid angle are too small. But it could have been produced in the wind from an accretion disk as matter flows into a black hole[9].

The temporal coincidence of SN2003dh with GRB030329 rules out the supranova model[23] for this event. Neither the jet itself, nor any γ-rays it made, could escape until the supernova had expanded for several months. The presence of large quantities of $^{56}$Ni necessary to account for the light curve may also be problematic in the magnetar model[24] where there is no accretion disk. Producing the $^{56}$Ni in a shock would require that the pulsar's energy be distributed to a large mass, which is inconsistent with producing a relativistic jet with almost no mass. Our observations thus support the collapsar model[8,9] for GRBs.

**Acknowledgements** We thank F. Patat for discussions. This paper is based on observations collected by the Gamma-Ray Burst Collaboration at ESO (GRACE) at the European Southern Observatory, Paranal, Chile. We thank the ESO staff at the Paranal Observatory, in particular N. Ageorges, P. Gandhi, S. Hubrig, R. Johnson, C. Ledoux, K. O'Brien, R. Scarpa, T. Szeifert and L. Vanzi, for their help in securing the service mode data reported here. We acknowledge benefits from collaboration within the EU FP5 Research Training Network "Gamma-Ray Bursts: An Enigma and a Tool". This work was also supported by the Danish Natural Science Research Council (SNF). J.P.U.F and K.P. acknowledge support from the Carlsberg Foundation.

**Competing interests statement** The authors declare that they have no competing financial interests.

**Correspondence** and requests for materials should be addressed to J.H. (e-mail: jens@astro.ku.dk).

**Table 1 ESO VLT optical spectroscopy of GRB030329/SN2003dh**

| Date | Instrument | Grism/Filter | Int. Time | Seeing | V magnitude |
|------|------------|--------------|-----------|--------|-------------|
| (2003 UT) | | | (s) | (FWHM in ") | |
| Apr 3.09 | FORS1 | V | 60 | 1.10 | $18.20 \pm 0.03$ |
| Apr 3.10 | FORS1 | 300V+10/GG435 | 1,800 | 0.6–1.0 | $20.5^{+\infty}_{-0.75}$ |
| Apr 8.12 | FORS1 | V | 60 | 0.99 | $19.30 \pm 0.03$ |
| Apr 8.13 | FORS1 | 300V+10/GG435 | 1,800 | 0.7–1.3 | $20.54 \pm 0.2$ |
| Apr 10.02 | FORS2 | V | 60 | 1.02 | $19.65 \pm 0.03$ |
| Apr 10.04 | FORS2 | 300V+20/GG375 | 3,600 | 0.8–1.8 | $20.25 \pm 0.15$ |
| Apr 10.10 | FORS2 | 300I+21/OG590 | 5,400 | 1.4–1.8 | |
| Apr 16.99 | FORS2 | V | 60 | 1.12 | $20.16 \pm 0.03$ |
| Apr 17.01 | FORS2 | 300V+20/GG375 | 900 | 0.67–0.81 | $20.38 \pm 0.2$ |
| Apr 21.99 | FORS2 | V | 60 | 1.01 | $20.53 \pm 0.03$ |
| Apr 22.00 | FORS2 | 300V+20/GG375 | 900 | 0.60–1.13 | $20.95 \pm 0.2$ |
| Apr 30.99 | FORS2 | V | 60 | 0.70 | $21.16 \pm 0.05$ |
| May 1.01 | FORS2 | 300V+20/GG375 | 1,800 | 0.44–0.68 | $21.72 \pm 0.2$ |
| May 1.03 | FORS2 | 300I+21/OG590 | 1,800 | 0.63–0.81 | |

The epoch of GRB030329 is Mar 29.4842, 2003 UT. The observations were obtained with FORS1 and FORS2 at ESO's 8.2-m Antu and Yepun telescopes at Paranal Observatory, Chile. At all epochs the observations were divided into 600-s or 300-s exposures to facilitate cosmic-ray rejection. A 1.3" wide slit was used, oriented so as to include a nearby star at position angle 123.6°. The data were reduced using IRAF in a standard manner. Flux calibration was performed using a spectrophotometric standard star and absolute flux calibration was done using the simultaneous V-band photometry. On the basis of the spectra of the nearby star we estimate that relative flux calibration errors are less than $\pm$ 5 %. The broad-band V magnitudes are for the combined optical afterglow and supernova. The photometric zero point is based on the secondary standard star USNO U1050_06350247 for which we use V = 16.84 (ref. 25). The spectroscopic magnitudes (grism) refer to the derived V magnitudes of the supernova following the decomposition (see Fig. 2). We used the series of fits and decompositions obtained with different templates to estimate the uncertainties in the reported magnitudes and folded in the potential large-scale flux calibration errors. No corrections for foreground reddening ($E(B - V)$ = 0.025, ref. 26) were applied.

**Table 2 Emission-line properties of the host galaxy of GRB030329**

| ID (2003 UT) | Wavelength (Å) | Redshift | Flux ($10^{-17}$ erg cm$^{-2}$ s$^{-1}$) | Luminosity ($10^{39}$ erg s$^{-1}$) | SFR M$_\odot$ yr$^{-1}$ |
|---|---|---|---|---|---|
| [O II] $\lambda 3{,}727$ | $4{,}355.5 \pm 2$ | $0.1685 \pm 0.0005$ | $17 \pm 2$ | $13 \pm 2$ | $0.18 \pm 0.06$ |
| [Ne III] $\lambda 3{,}869$ | $4{,}520.0 \pm 2$ | $0.1682 \pm 0.0004$ | $3.5 \pm 0.7$ | $2.7 \pm 0.5$ | |
| He I $\lambda 3{,}889$ + H8 | $4{,}542.6 \pm 2$ | $0.1681 \pm 0.0004$ | $1.1 \pm 0.4$ | $0.9 \pm 0.3$ | |
| [Ne III] $\lambda 3{,}968$ + H$\epsilon$ | $4{,}638.0 \pm 2$ | $0.1686 \pm 0.0004$ | $2.4 \pm 0.6$ | $1.9 \pm 0.5$ | |
| H$\delta$ $\lambda 4{,}102$ | $4{,}792.0 \pm 2$ | $0.1683 \pm 0.0004$ | $1.1 \pm 0.4$ | $0.9 \pm 0.3$ | |
| H$\gamma$ $\lambda 4{,}340$ | $5{,}071.6 \pm 2$ | $0.1684 \pm 0.0004$ | $4.5 \pm 0.8$ | $3.6 \pm 0.6$ | |
| H$\beta$ $\lambda 4{,}861$ | $5{,}679.3 \pm 2$ | $0.1683 \pm 0.0004$ | $10 \pm 2$ | $8.1 \pm 2$ | |
| [O III] $\lambda 4{,}959$ | $5{,}793.5 \pm 2$ | $0.1683 \pm 0.0004$ | $14 \pm 2$ | $11 \pm 2$ | |
| [O III] $\lambda 5{,}007$ | $5{,}849.1 \pm 2$ | $0.1682 \pm 0.0004$ | $40 \pm 5$ | $31 \pm 4$ | |
| H$\alpha$ $\lambda 6{,}563$* | $7{,}667.1 \pm 2$ | $0.1683 \pm 0.0003$ | $28 \pm 3$ | $22 \pm 2$ | $0.16 \pm 0.05$ |
| [N II] $\lambda 6{,}583$ | $7{,}693.3 \pm 2$ | $0.1686 \pm 0.0003$ | $3.3 \pm 0.7$ | $2.6 \pm 0.5$ | |

The emission line measurements refer to the 1 May spectra (see Table 1). The reported wavelength and redshift uncertainties are dominated by systematic wavelength calibration errors. An accurate estimate of the redshift of the host galaxy was derived using the strong H$\beta$ and [O III] lines by fixing the wavelength of the nearby [O I] skyline to 5,577.3 Å. The resulting redshift is $0.1686 \pm 0.0001$. The properties of the host galaxy of GRB030329 are discussed in Fig. 1. Also shown is the star formation rate of the (probably large) fraction of the host galaxy covered by the slit, based on the calibrations given in ref. 27.

* Measurements based on the H$\alpha$ line are marginally affected by telluric lines.

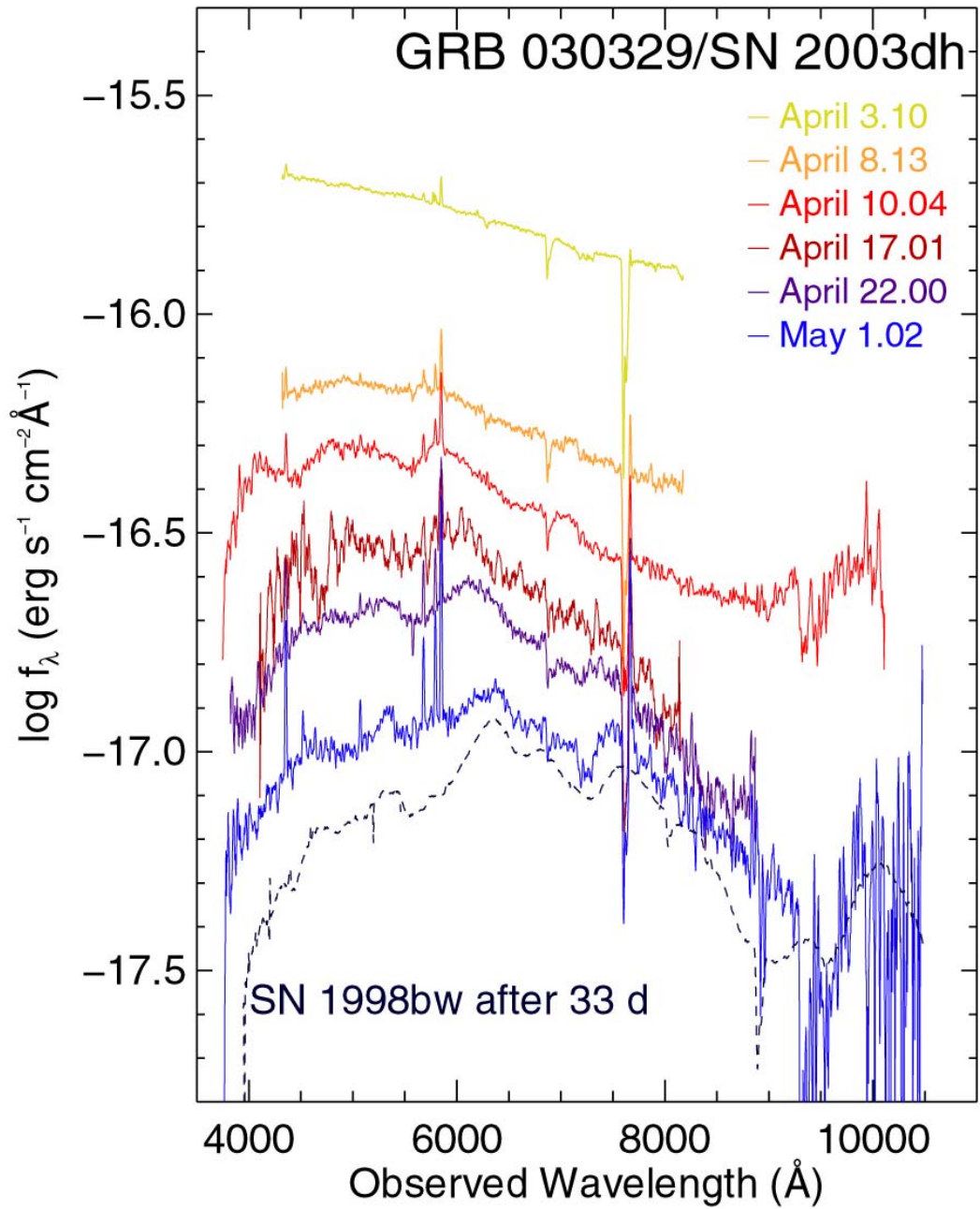

**Figure 1** Spectral evolution of the combined optical flux density, $f_\lambda$, of the afterglow of GRB030329, the associated SN2003dh, and its host galaxy. The details of the observations are given in Table 1. The upper spectrum is rather well fitted by a power law, as usually seen in afterglow spectra. The middle spectra show clear deviations from a power law, similar to SN1998bw at the same phase. The lower spectra, dominated by SN2003dh, reveal the supernova signatures. For comparison, the spectrum of SN1998bw after 33 days[21] is shown (dashed line) shifted to the GRB030329 redshift. All SN2003dh spectra are presented in observed wavelengths, and no reddening correction has been applied. Telluric absorption lines have also been left in the spectra. The region above 9,000 Å is hampered by sky-line emission, but we tentatively identify a broad feature centered at 10,000 Å that could be due to the supernova. At all epochs we identify emission lines of [O II] $\lambda 3{,}727$, H$\beta$, [O III] $\lambda 4{,}959$ and $\lambda 5{,}007$ and H$\alpha$ (Table 2), most probably from the host galaxy. At the last epoch (1 May) we also identify [Ne III] $\lambda 3{,}869$, H$\delta$, H$\gamma$, [N II] $\lambda 6{,}583$ as well as blended lines of He I $\lambda 3{,}889$ + H8 and [Ne III] $\lambda 3{,}968$ + H$\epsilon$. From the strengths of the Balmer lines we infer that the extinction in the host galaxy is small. The metallicity based on the [O II], [O III], and H$\beta$ fluxes is [O/H] = – 1.0 (ref. 28). The derived star-formation rate is about 0.2 M$_\odot$ yr$^{-1}$ using either [O II] or H$\alpha$ (Table 2). Using the 3$\sigma$ upper limit of R > 22.5 derived for the magnitude of the host from archival data[29], we conclude that the equivalent widths of the emission lines are very high. The host galaxy is thus an actively star-forming, low-metallicity, dwarf galaxy and appears to be qualitatively very similar to the host galaxy of GRB980425/SN1998bw (ref. 30).

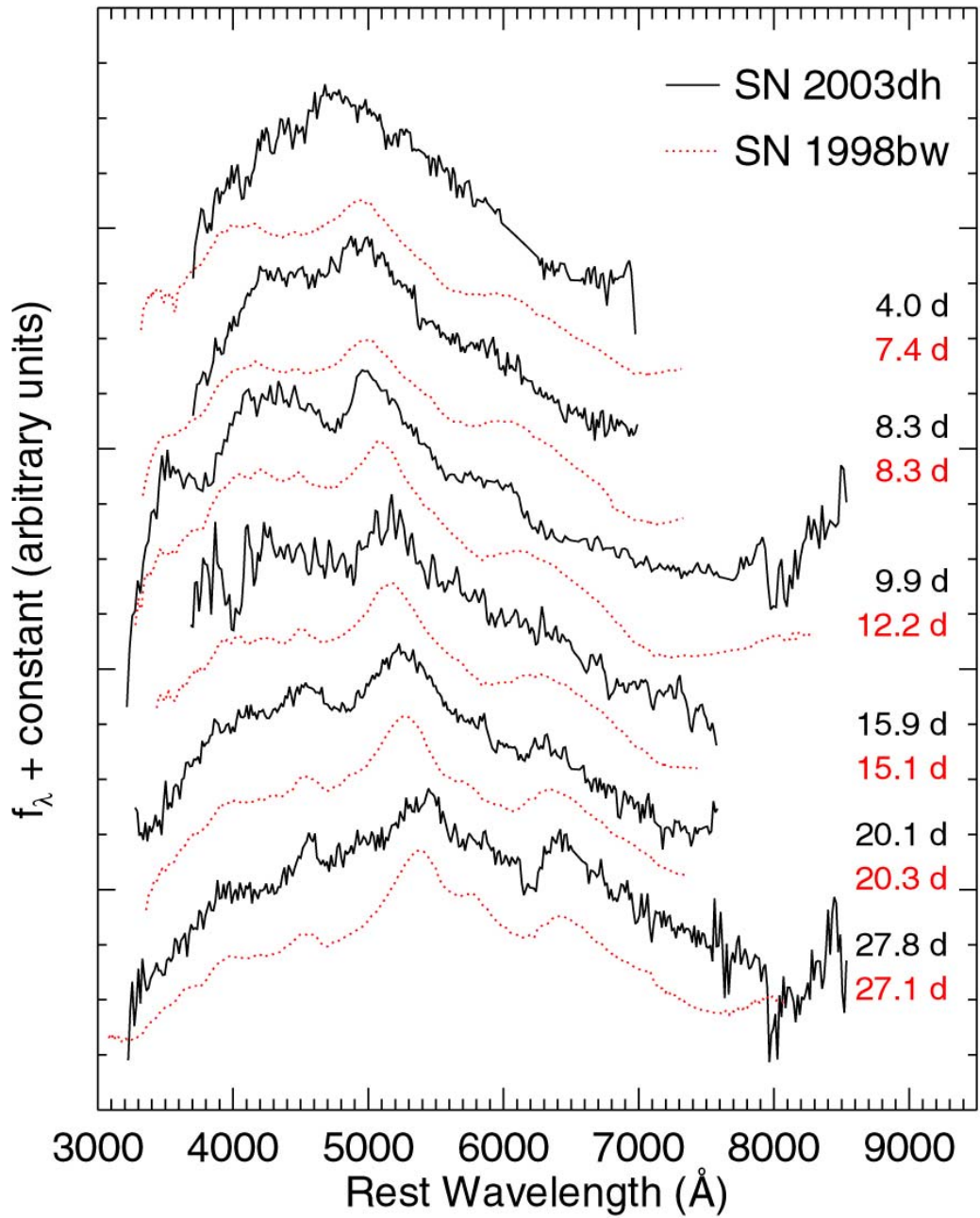

**Figure 2** Comparison of the spectral evolution of SN2003dh and SN1998bw.
Solid lines indicate spectra of SN2003dh obtained through decomposition as
described below. Dotted red lines indicate spectra of SN1998bw taken at similar
epochs[15]. All spectra are plotted as a function of restframe wavelength and
labeled with the time since the GRB, corrected for the cosmological time dilation

(1+*z*); the relative vertical offsets are arbitrary. The SN2003dh spectra are binned to 15-Å bins, the emission lines from the host galaxy have been removed, and where necessary the strong telluric absorption features at 6,800 Å, 7,200–7,400 Å, and 7,600 Å have been interpolated. The strong absorption 'edge' at 6,150 Å (7,200 Å observed) is probably due to telluric absorption. The spectral decompositions were performed as follows. While the host galaxy has strong emission lines, its continuum flux upper limit is negligible at early epochs and significantly less than the total flux at the later epochs. The contribution from the host galaxy was therefore accounted for by simply removing the emission lines. Model spectra were constructed as a sum of a power law ($f_\lambda \propto \lambda^{-(\beta+2)}$) and a scaled version of one of the SN1998bw template spectra[15]. For each template, or section thereof, a least squares fit was obtained through fitting of the three parameters: power-law index $\beta$, amplitude of afterglow, and amplitude of supernova. In most cases the best fitting index was found to be $\beta \sim -1.2 \pm 0.05$ which was adopted throughout. We note, however, that the resulting overall spectral shape of the supernova contribution does not depend on the adopted power law index or template spectrum. The striking similarity between the spectra of these supernovae is clearly seen. The spectral peak wavelength, for both supernovae, is shifting towards the red. The shift is on average ~25 Å per day for SN2003dh, which is similar to the evolution of the early spectra of SN1998bw. The cause of this shift is the growing opacity in the absorption blue-wards of 4,900 Å (rest wavelength). The spectral decompositions provide the fraction of the total flux in the V band that is due to the supernova. The resulting SN2003dh V magnitudes are reported in Table 1 and plotted in Fig. 3.

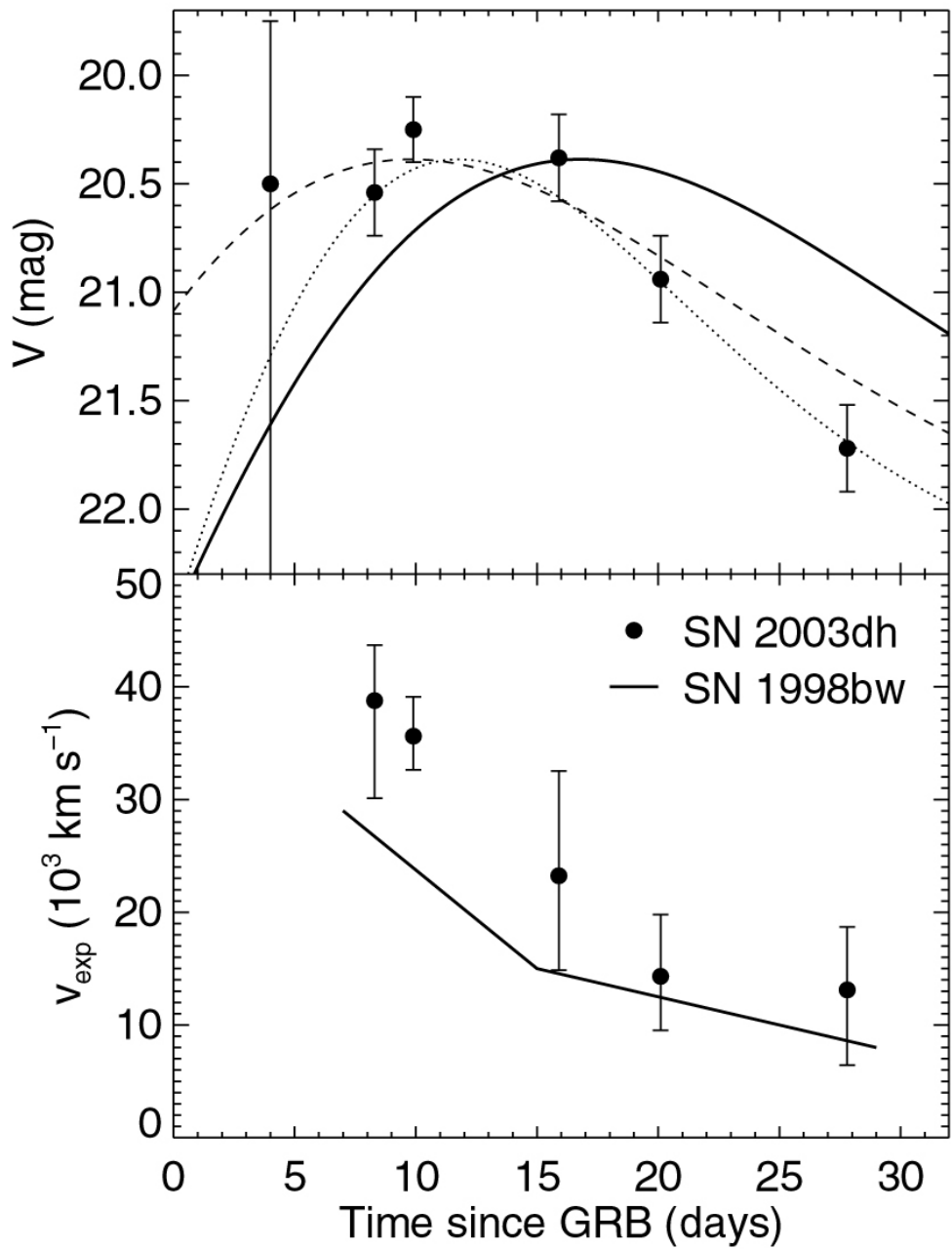

**Figure 3** Light curves and expansion velocities of SN2003dh and SN1998bw.

Upper panel **a**, Light curves. Filled circles, spectroscopic V magnitudes of

SN2003dh (Table 1) as a function of time (restframe) since GRB030329. Solid

line, the brightness of SN1998bw[1] as it would have appeared in the V band at *z*

= 0.1685 as a function of time (restframe) since GRB980425. Dashed line, as for the solid line but shifted 7 days earlier. Such an evolution may be expected if the supernova exploded 7 days before the GRB. For SN2003dh, this however is inconsistent with its spectral evolution (Fig. 2). Dotted line, as for solid line, but evolution speeded up by multiplying time by 0.7. A faster rise and decay may be expected in asymmetric models in which an oblate supernova is seen pole-on. We assumed 0.20 mag (refs 1, 15) extinction for SN1998bw and none for SN2003dh. Bottom panel, **b**, Si II λ6,355 expansion velocities as a function of time (restframe). Filled circles, SN2003dh; solid line, SN1998bw. The SN2003dh values are our best estimates based on the decomposed spectra (Fig. 2). We caution that in some cases these values are very uncertain due to other features in the spectra around the expected minimum. The solid line shows the trend for SN1998bw based on the data points in ref. 15. The consistent decaying trend in the ejecta velocity of SN2003dh, together with its very high initial value, indicate that we observed very close to the supernova explosion.